# Hitting the ground running:
# Computational physics education to prepare students for computational physics research

Amy Lisa Graves
Swarthmore College Dept. of Physics and Astronomy
Swarthmore, PA 19081

Adam Light
Colorado College Dept. of Physics
Colorado Springs, CO 80903

**Keywords:** Computational Physics, Undergraduate Research, Physics Curriculum, Research Skills

**Abstract:**
Momentum exists in the physics community for integrating computation into the undergraduate curriculum. One of many benefits would be preparation for computational research. Our investigation poses the question of which computational skills might be best learned in the curriculum (prior to research) versus during research. Based on a survey of computational physicists, we present evidence that many relevant skills are developed naturally in a research context while others stand out as best learned in advance.

1. Introduction

The benefits of cutting-edge research with undergraduates are undeniable, in computational physics (CP) or any other discipline. Research brings many benefits to students, with special ones attached to students who are early in their career, women and/or members of underrepresented minority groups[1]. The purpose of this paper is to ask how to best prepare a student for CP research (CPR) which in the U.S. typically happens during the summer, whether on the home campus, or elsewhere via a program like the National Science Foundation Research Experience for Undergraduates (NSF REU). Arriving prepared at a CPR lab, whether at home or away, implies a greater chance of thriving: comfort with the work, higher self-concept as a researcher, faster progress, and more likelihood of eventual co-authorship on conference posters, talks and refereed papers. Whether in a lab, or as a culminating experience in an undergraduate course or capstone, research is "increasingly important for our students' careers" (Editors' Introduction, Ref. [2]).

We ask about preparation as advocates not just for students, but faculty members as well. Faculty at principally undergraduate institutions (PUIs) typically have only undergraduates as their junior partners in research. A track record of cutting-edge research is essential for the careers of most PUI faculty members, making it imperative that there is authentic research progress in an 8- to 10-week summer. The current paper originates in a 2019 talk (found at meetings.aps.org/Meeting/MAR19/Session/F22.3). To our knowledge, we are among the first to publically frame the question of how undergraduate students can "hit the ground running" in CPR.

In 2011, the AAPT recommended that since computation is integral to the modern practice of physics, "every physics and astronomy department provide its majors and potential majors with appropriate instruction in computational physics".  In 2016 the AAPT Undergraduate Curriculum Task Force (UCTF), produced an informative rationale for fundamental learning of computer-based tools, and modes of computational thinking and doing science[3,4,5].  This report identifies challenges departments will face in implementing the suggestions, and resources for faculty. Skills are broken into three categories that build upon one another: fundamental, technical, and computational physics-related. In our study (see Section 3), we attempt to segregate basic, fundamental, and more advanced skills as well as workplace skills like finding information, collaborating with others, and project management (also key in a graduate school setting)[3].

In addition to designing new courses and repurposing existing ones, experts recommend leveraging "opportunities for computational ... research projects"[5] and fostering opportunities where "undergraduate research is valued and supported"[4].  In short, CP skills are recognized as the fruit of undergraduate CPR.  The premise of this paper is that the reverse is also true. Computational skills are needed to engage in computational research, which builds further skills and advances research progress.

## 2. Computation in the undergraduate curriculum

We make a distinction between computational physics education (teaching students more physics, more effectively with computers as a vehicle) and education in the discipline of computational physics[2, 6, 7].   (Utilizing simulations such as Physlets, or PhET simulations from U. of Colorado, though excellent pedagogy, doesn't count as the latter.) Further, there is the well-recognized split between a "scientific computing approach" vs. a "computational science" approach to CP education (N. Chonacky and D. Winch's article,  Ref. [2]). Is it better to teach the theory of a $n^{th}$ order symplectic integrator; or ask students to find one in a repository on the Web, download and utilize it; understanding its limitations and interpreting the results via the theory of dynamical systems? The "understand everything from the ground up" physicists' mindset predisposes toward the former.  Yet the latter may be better preparation for CPR in dynamical systems theory, and/or the world of employment beyond college.

A decade ago there were only five U.S. undergraduate degree programs in CP[R.H. Landau in Ref. (2)].  An updated report (PICUP and the AIP Statistical Research Center) surveying all U.S. physics departments is forthcoming, but a rough guess can be made that this number has increased, but by less than an order of magnitude.  Somewhere between 25% and 50% of departments teach computation at the introductory or advanced level[8, 9].  The state-of-the art roughly a decade ago was ably reviewed in special issues of CISE (Volume 8, 2006) and the American Journal of Physics [2]. Valuable curricular resources in the latter include a pedagogical resource letter(R.H. Landau) and a guidelines paper based on two years of meetings and surveys with experts and stakeholders (N. Chonacky and D. Winch). That era might be construed as the "2nd generation" of CP education books and courses[8], in which pure numerical analysis texts gave way to being more oriented toward physics, with support for projects.  A clue that CP courses are already structured to benefit CPR is that the faculty responsible for CP curricular innovation have, then and now, been found to be those engaged in CPR[8,9].

Difficulties in initiating/sustaining a CP curriculum have been recognized.  These "systemic forces" include[5,10]
- it's time consuming (instructors must prepare; time for traditional topics is lost)
- it disrupts the status quo of courses offered
- it requires new background (for instructor and student)
- it requires more than one faculty member - a community
- we lack research, from the PER community or elsewhere, on effective teaching of CP

Hopefully, potential instructors are reassured that they are not alone in encountering these difficulties. They are discussed in light of a recent PICUP-sponsored faculty workshop in Ref. [10]. On the brighter side, there are many resources available for curricular support and change, including:
- campus ITS departments, a number of whom are already offering trainings in Git, command line tools, the XSEDE HPC framework, and more.
- software bootcamps, in person or online through organizations like software-carpentry.org, datacamp.com, and Lynda.com
- comPADRE.org, which is a general AAPT-supported materials repository and includes resources specific to CP like Open Source Physics (OSP), and the PICUP project.  PICUP in particular is well aligned with the goals expressed here and facilitates the "organic" use of computation in traditional courses.
- shodor.org, which features the *Journal of Computational Science Education* and hosts the Computational Science Education Reference Desk (CSERD) and the National Computational Science Institute (NCSI).  These resources provide curriculum and workshops for faculty seeking to teach computational science.

Vis a vis preparation for CPR, physics might gain insight from materials science and engineering (MSE). Computational MSE is now an expected departmental research thrust.  In 2005 most of the top 20 U.S. MSE departments were planning to introduce computational undergraduate courses and 43% already had, by 2009 the majority of departments had done this[11]. Courses like these tend to teach research-caliber tools; such as industry-standard codes for density functional theory, molecular dynamics, finite element modelling, and thermodynamic phase diagrams; and a notebook environment for calculation/visualization[12]. This is consistent with the fact that most MSE faculties prioritize courses with a computational science approach; for example, only a small minority of MSE faculty wanted students to learn to write code and create new tools[11]. Stated aims of computational MSE courses are, as in physics, deepened understanding of traditional topics. However, it is clear that MSE courses which teach undergraduates to become confident modelers with state-of-the-art software satisfy the need to hit the ground running in undergraduate research, graduate school and the workplace.

The PER literature reveals how active learning precepts[13] translate readily to the typical hands-on, team-based teaching of CP in which all stages of manipulating code, data and results ask students to construct their own knowledge of physics. Since "Most of our knowledge of how to teach ... CP is anecdotal."[5] and, as mentioned above, few PER studies have concerned CP education[8,14], there exist at present "significant opportunity for the PER community to support the development of materials, teaching practices and assessment strategies."[9]. A motivation for the current paper is to explore the opportunity for educational experts and CPR stakeholders to factor in CPR readiness. We wonder which computational skills are best acquired outside of and brought to the lab versus learned (and enhanced) in the lab, and what practices make for inclusive CP education.

**3. The survey: What skills are needed at what stage?**

Respondents who do CP research with students (see Section 4) were asked "Please assess the type of experience undergraduate students need before or during their time with you in order to be effective, productive researchers in your computational physics lab."

Categories A and H below asked for text answers. For B-G, respondents were asked to check one or more of four boxes:
- Ideally, students have prior experience with these
- Students get focused training on these
- Students learn these in natural course of research with me
- This skill is unnecessary for my work

Survey categories were as follows:

A. Field of Physics in which I do computation-based research

B. Algorithmic thinking skills
   1. Converting a problem to a step-by-step procedure amenable to coding
C. Basic coding skills and knowledge
   1. Programming in some language at "hello, world!" level
   2. Calculation/visualization environment (e.g. Mathematica, Jupyter notebook, Matlab)
   3. Numerical arithmetic (e.g. machine precision, data types)
   4. Flow control (loops, conditionals, etc.)
   5. Scoping rules particular to a programming language
   6. Selecting and manipulating data structures
D. Software engineering
   1. Whiteboarding
   2. Pseudocoding & diagramming

3. Software lifecycle best practices (design-code-implement-validate-utilize-archive)
   *(Note: This answer could be "no", despite a "yes" to the step of testing code.)*
4. Keeping a TODO list
5. Documenting code and keeping documentation current
6. camelType and other strong typing
7. Interfacing homegrown code with external libraries
8. Object oriented design principles
9. Fluency in an object oriented language
10. Parsing/understanding legacy or inherited code
11. Extending functionality of existing code
12. Version control
13. Creating modular, well-structured code easily passed on to other students
14. Using an IDE

E. Scientific project and data management
   1. Keeping a lab notebook or electronic log
   2. In situ documentation (README files, etc.)
   3. Recording input parameters when codes are run in production mode
   4. Recording version of code used to produce any set of data
   5. Naming conventions of output files
   6. Backing up and knowing how to restore data/codes
   7. Sharing code effectively with others
   8. Proactive communication of results and issues
   9. Collaborative mindset
F. Fundamental scientific computing skills
   1. Running an executable code one is given
   2. Asking people for help with running or writing code
   3. Reading code documentation
   4. Googling to answer coding questions
   5. Locating existing code and libraries on web-based servers
   6. Terminal / unix shell programming
   7. Unix shell scripting
   8. Accessing and utilizing remote platforms (ssh, scp, etc.)
   9. Submitting jobs to a queue
   10. Using a terminal-based text editor
   11. Compiling and making
   12. Reading formatted data
   13. Writing formatted data
   14. Understanding compile-time and runtime errors
   15. Visualizing line data (plots)
   16. Visualizing image data (2D)
   17. Visualizing volumetric data (3D)
   18. Creating animations
G. More advanced computational skills

1. High performance computing
2. Profiling code and optimizing execution
3. Specialized hardware (GPUs, etc.)
4. Using a debugging environment

H. What have we forgotten to ask about?  What do your students need to "hit the ground running" in your computational research?

## 4. Forty-six computational physicists respond to survey

The survey of Section 3 was sent to a total of 105 physicists and astronomy faculty, of which 35 and 70 taught at colleges (PUIs) and universities respectively.  This corresponded to 27 and 29 unique institutions.  We began with the *U.S. News and World Report* 2019 list of twenty top U.S. colleges and universities. Websites for their Physics Departments were scrutinized for computational research that involved undergraduates. Identified faculty members received email solicitations and a link to the survey.   In addition, we contacted 19 faculty members known to us for CPR with undergraduates.  The overall response rate was 44% (substantially higher for colleges than universities.) The fields of respondents were diverse: atomic physics, astronomy, biophysics, condensed matter, cosmology, fluids, gravitation, materials physics, nonlinear dynamics, nuclear, plasma, soft matter, statistical physics, and quantum: field theory, information theory, mechanics and optics.  Compelling questions about differences between PUIs and universities or between different subfields must be left for a future study.

A comprehensive look at the survey data indicates that some skills are better suited to learn "on the job" than others.  The scatter plot in Figure 1 captures four dimensions of the data.  Each circle corresponds to a labeled skill (see Section 3). We plot skills for which prior knowledge is desired vs. those acquired naturally or via focused training in the lab. The circle radius indicates fraction of respondents for which the skill is needed. Color shows fraction who assert the skill is acquired naturally during research.

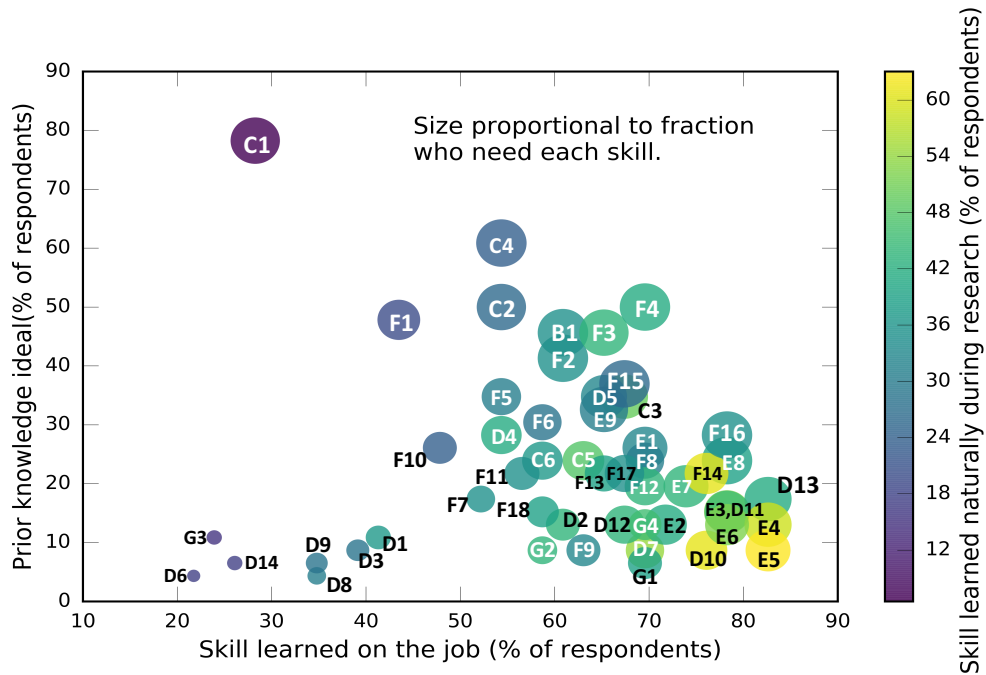

*Figure 1:* Summary of computational skill survey data. Each circle corresponds to one skill/survey question (see Section 3). Circle radius is scaled by the fraction of respondents whose students needed that skill; the largest circles (e.g. C4) represent skills needed by 100% of respondents

A coarse interpretation of this plot is that the skills in the upper left are perhaps necessary to teach in the curriculum. Skills on the lower right can (perhaps should) be learned as needed during the course of the research. The need for basic programming skills [C1] to be developed ahead of time is particularly clear in this representation. 98% of respondents found it necessary, most agreed that students should have prior experience before beginning research, and few believed it develops naturally during the course of research. Of the 52 skills surveyed, six of them had *all* respondents identify them as necessary:

- [B1] Converting a problem to a step-by-step procedure amenable to coding
- [C4] Flow control (loops, conditionals, etc.)
- [E8] Proactive communication of results and issues
- [F2] Asking people for help with running or writing code
- [F4] Googling to answer coding questions
- [F15] Visualizing line data (plots)
- [F16] Visualizing image data (2D)

The type of experience respondents identified as necessary for students to have with these skills was not uniform, however. Responses for [B1] and [E8] above, as well as [C1] (basic programming) and [E4] (code version/output association) are compared in the pie charts of Figure 2. The vast majority of respondents valued these skills, and they cover a range of the tradeoff between prior knowledge and natural skill development.

An overwhelming number (80%) of respondents emphasized a need for prior experience with programming [C1]. Respondents split fairly evenly on whether the skill [B1] of translating a problem into procedural code was a best learned ahead of time or naturally in the course of research, with another 15% advocating focused training. A third of respondents indicated that students should come in with [E8], ability to proactively communicate results and issues. The mundane but vital skill, valued by 93% of respondents, of matching code versions to data produced [E4] was learned primarily on the job.

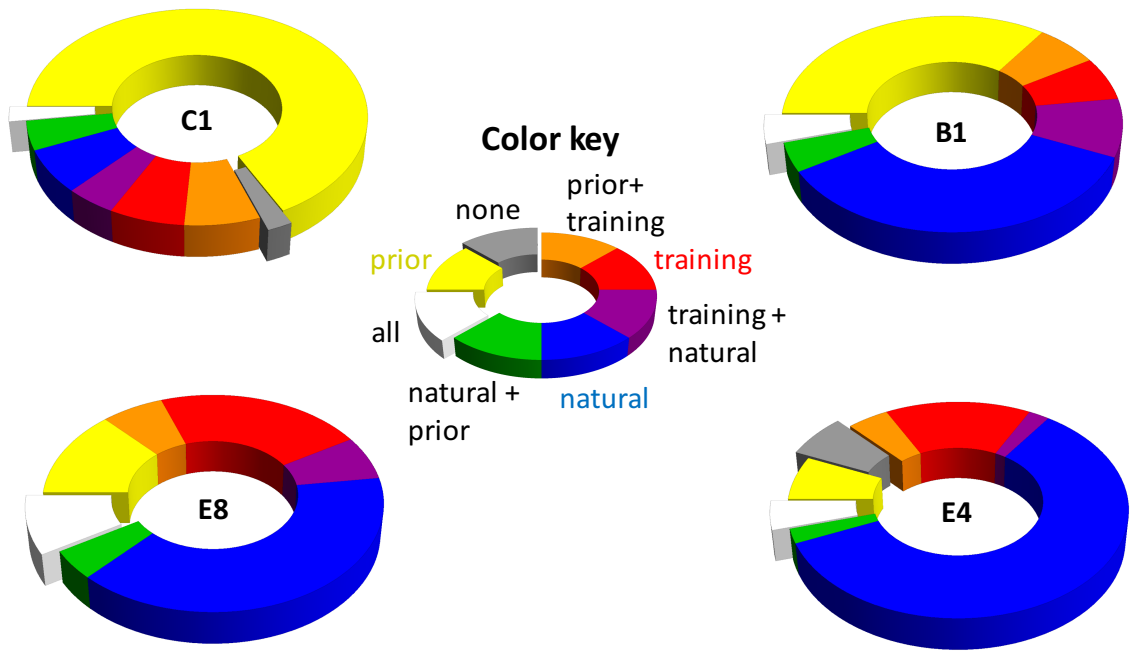

*Figure 2:* Breakdown of responses for particular skills discussed in text. Primary responses: blue, red and yellow; respondents choosing two primary responses: orange, purple and green; choosing all three: white; categorizing skill as unnecessary: gray.

**5. Do survey results map to typical CP curricula?**

It is a rebuttable presumption that a physics course in computation will instill many, but not all, skills ideally needed prior to CPR. For example, consider the following four excellent undergraduate courses:

• Sharma's freshman-level course for potential STEM majors at Wagner College[15] exposes students to work flows in producing, analyzing and visualizing technical data. With no computer science prerequisites, it distances itself from being a programming course. Mathematica, open-source molecular dynamics, and molecular visualization packages are taught. As with MSE computing courses discussed in Section 2, the emphasis is on using high level platforms for insight into high level questions. Integral to the course are manipulating and visualizing data structures in Mathematica, from molecular mechanics simulation, and from Protein Data Bank files. Some of the substantial final projects have a level of innovative exploration which qualifies as computational research.

• Caballero and Pollock's PER-informed, Mathematica-based, intermediate classical mechanics course at U. Colorado at Boulder seeks both to deepen an understanding of traditional topics and to prepare students for professional physics research[14]. It is pragmatic in terms of student and faculty time. For example, theory of error analysis is replaced with common sense checks like running limiting cases, a "critical part of the modeling cycle"(Editors' Introduction, Ref. [2]). .

• Bryn Mawr College has developed an extensive set of materials for both instructors and students [16]. Funded by TIDES, a Diversity/Equity initiative of the AACU and its Project Kaleidoscope (PKAL), a team developed 14 iPython/Jupyter notebook-based modules which embed computational lessons in physics/engineering and social context.  Students' electronic portfolios contain computing work plus profiles of computational scientists and reflections on how what was learned intersects with personal goals. It is clear, for example from the National Center for Science and Engineering Statistics *Women, Minorities, and Persons with Disabilities in Science and Engineering: 2019*. Special Report NSF 19-304, that the physical and computer sciences are fields that persistently skewed away from women and people of color.
Thus, goals of equity and inclusion are singularly valuable.

	• Swarthmore College's computational lab has undergone many incarnations in 30 years - using Basic, IDL, Matlab, Mathematica, and (currently) with iPython/Jupyter notebooks. Currently, the course is 7 weeks long and preceded by a recommended workshop, run by the Software Carpentry Foundation of San Francisco, CA and covering terminal editing, Unix shell, and use of a local Github for version control.  Topics in numerical analysis are integrated with physical applications, with a final, synthetic exercise on LIGO gravitational wave data.

	In Table I we've noted which surveyed skills are listed as explicit course goals; with the caveat that names of categories B-G are not rigorous identifiers, and are open to interpretation.

|  | Wagner | Colorado | Bryn Mawr | Swarthmore |
| --- | --- | --- | --- | --- |
| B. Algorithmic thinking | 1 | 1 | 1 | 1 |
| C. Basic coding | 1-6 | 2, 3 | 1-4, 6 | 1,2,4-6 |
| D. Software engineering | 6, 11 | None | 2, 5, 7, 8, 11, 13, 14 | 5,7,10,11,13 |
| E. Scientific Programming | 1, 5-9 | 1, 8, 9 | 1, 9 | None |
| F. Fundamental computing | 1-4, 12, 13, 15-18 | 1, 2, 4, 15-18 | 1-6, 10, 12-16 | 1-4, 12-16 |
| G. More advanced | None | None | 2, 4 | None |

*Table I. Skills detailed in Section 3 which were also course goals.*

*Notes: i. Absence of D3 is not indicative of a lack of "validation", since several courses included testing of known cases. ii. D14 may refer to the Jupyter environment, not a separate IDE.*

We invite readers to cross-compare Table I with CPR expert priorities displayed in Figures 1 and 2. Also consider the handful of skills which *none* of the four courses had as an instructional goal: D1, D3, D4, D9, D12; E2-4; F7-9, F11; G1 and G3. These omissions are quite consistent with Figure 1. The least consistent was [D4] keeping a TODO list; even so, less than 30% of respondents wanted prior experience, around 50% said it would be learned on the job, and 20% it was not needed at all. Understandably, some items are absent because they are outside the technical scope of course tools (e.g. terminal editing using a high level language, HPC). Even so, students did feel the lack of some of these, such as the ability to debug their programs.

Outside the technical reach of many course are certain skills in which roughly 20% of survey respondents wished prior training: [F7] Unix shell scripting, [F8] connecting with remote platforms, [F11] compiling and making, and [E2-4] writing READMEs and keeping track of versions and input parameters. (Item [F6] terminal and Unix basics, was equally desirous for CPR, and a goal of only one of the four courses.) On the other hand, skill [D4] Keeping a TODO list, was one in which roughly 30% of respondents wanted prior preparation. This pre-professional skill worth teaching would not stretch the technical scope of any course.

## 6. Advice and innovation from folks doing CPR

One CP course which deserves careful mention[17] is focused on CPR readiness and rooted in expert advice. Burke and Atherton, for their 2015 Tufts University course (mostly undergraduates at roughly a sophomore level) modified their initial list of competencies based on interviews with faculty and postdocs active in CPR. Their evidence-based model of expert problem solving was less linear than their original model (Figure 3 below), and included meta-competencies seldom found in traditional undergraduate CP courses.

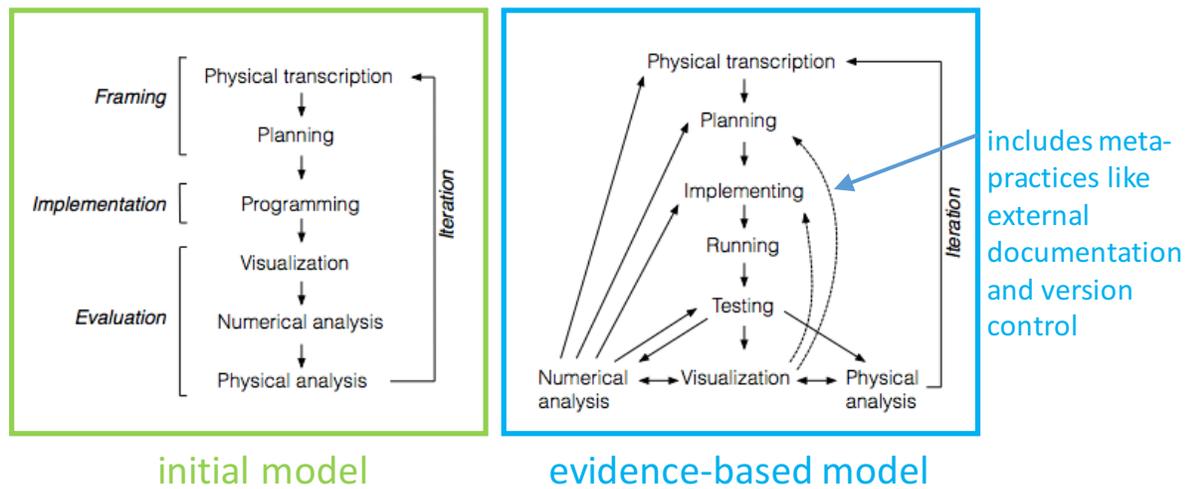

**Figure 3:** *Adapted from Figures 1 and 2 of Am. J. Phys. 85, 301 (2017), with the permission of the American Association of Physics Teachers.*

The learning goals, in a nutshell, were: (1) model a physical system (2) design-implement-validate code (3) understand the importance of numerical analysis (4) test hypotheses via visualization and other means (5) connect theory, experiment and simulation. A physics-content-driven, project-based course is a time-honored way to progressively introduce successively deeper knowledge of the various elements of CP, a fact in evidence in many widely-used CP textbooks. (See for example Rebbi's article in Ref. [2]; Gould, Tobachnik and Christian's Introduction to Computer Simulation Methods, 2006; or Landau, Päez, and Bordeianu's Computational Physics: Problem Solving with Python, 2015). It has also been noted that such courses are both consistent with the ethos of a liberal arts education and valuable preparation for sizeable research projects[18], to wit CPR, in the future. In Ref. [17], learning was implemented via short projects in classical, quantum and statistical physics, culminating in a final project of the student's choice. A subset of learning objectives were highlighted for each project; these included authentic research-related activities like reading papers, consulting Internet resources, working in groups, running code, testing performance, documenting, and version control. Mathematica and Python were chosen as the required programming languages, but students were allowed to use others for the final project, providing yet another pragmatic lesson: cross-comparing different languages in context. Course evaluations indicated success, but pointed to constructive changes such as additional scaffolded learning for students with different backgrounds, and better management of group dynamics.

Turning to our new survey data, suggestions offered in Item H (see Section 3) can be distilled into various categories. Some respondents

I. asserted that students should
- have programming and prerequisite mathematics
- be able to convert a problem to a stepwise procedure for coding (survey Item B1)
- know specific numerical techniques, e.g. FFT, nonlinear curve fitting, or linear algebra
- be able to refactor an existing code
- be able to create new or substantially extend algorithms
- know to test a code - even if it is via a "reality check" via a known case. (Some respondents work in environments which don't have debugging capabilities.)
- think in common-sense terms about algorithms and workflow. For example
  *"to structure and automate computations so that the output of one calculation can be efficiently passed as input to the next"*
- to have algorithmic thinking skills paired with an associated understanding of the underlying science
- be proactively communicative
- be interested, dedicated, and motivated

II. wondered how to teach their students
- to be *"fearless"* about trying what may not work, and persist until they succeed
- *"soft skills"* of computing like commenting/version control/writing READMEs
- skills that they themselves may not have learned, or don't employ to the extent that best-practices would dictate. As one said:
  *"My coding experience was mostly self-taught and I gradually learned the(se) lessons the hard way".*

Several respondents described a self-paced bootcamp experience which they have devised for students. For example:
*"I have them log into my computational cluster to work through some standard canned exercises ... students work in teams. Two students can be working on slightly different versions of the same problem ... That way, they have a community to talk to, but they have ownership over their own results."*

Finally, while several respondents struggled with the rarity of having students who hit the ground running; at least one (from an R1 university) was OK with this, expecting undergraduates to arrive needing instruction in very basic skills.

A final expert perspective comes from Caballero's coded notes from interviews with CPR professionals in academia and industry. Ref. [19] breaks skills into several useful categories, most of which appeared on our survey directly or were addressed in one or more free responses. The only category without representation was meta-knowledge of what a

computational solution can deliver.  Resonances between the skills in Ref. [19] and our survey data are:

- Knowledge of certain computational methods, both numerical and data-centered (free responses)
- Knowledge of specific physical concepts (free responses)
- Move fluidly between math, physics and computation-related ideas (free responses)
- Professional practices like planning/writing pseudocode and debugging (addressed directly)
- Data analysis and visualization (addressed directly)
- Computational thinking: breaking a complex task into steps which can be encoded (addressed directly)
- Meta-knowledge of what a computational solution can deliver (not addressed)
- Professional practices like documenting code and version control (addressed directly)

Interviewees in Ref. [19] also made a point echoed by several of our survey respondents: They were self-taught in professional CP practices. There was agreement, however, that this was not a situation that should carry forward into the education of the next generation of physicists.

## 7. Conclusions

CP is succinctly defined by the European Institute of Physics (IOP) website as "the science of using computers to assisting in the solution of physical problems, and to further physics research". The IOP's Computational Physics Group lists a remarkable array of computational physics subfields, suggesting it may be hopeless to identify a single type of worthwhile undergraduate preparation for all.  Nevertheless, the CPR and PER communities would benefit from coming together to classify types of skills and best teaching practices.  This issue is expressed by one survey respondent, who wishes they had an established program to bring students up to a needed level of preparedness *"instead of it being a case-by-case, ad hoc training situation"* which works for a small numbers of students, but not in perpetuity with large numbers of students in their research pipeline.  Another respondent remarks that one might similarly wish to prepare incoming *graduate students* with needed breadth of skills. Given their data-intensive field, this person remarks *"you unfortunately have to either teach them computation or physics.  As these skills are learned in different schools within universities ... I'm not aware of any undergraduate program that I think streamlines people to work with me."*

In conclusion, we hope that our current study is a "proof of principle" that experts and stakeholders can teach us which skills are needed, yet not easily mastered, during a brief stint doing undergraduate CPR. In this way, we inform ourselves how to structure CP curricula to help students hit the ground running.

# 8. Acknowledgements

We are grateful to Joan Adler and Rubin Landau for the opportunity to present our work. We thank the colleagues who generously responded to our lengthy survey. Additional thanks to Tony Atherton, Lorena Barba, Daniel Caballero, Catherine Crouch, Robert Hilborn, Paul Jacobs, Michael Lerner, Mark Matlin, Robert Panoff, Arjendu Pattanayak, Andrew Ruether, Arunkumar Sharma, and Christopher Tunnell. We gratefully acknowledge funding from the Office of the Provost and (AG) a James Michener Faculty Fellowship of Swarthmore College.

**Author Biographies:**

Amy Graves (formerly Amy Bug) is the Walter Kemp Professor in the Natural Sciences at Swarthmore College. Educated in mathematics and physics at Williams College and in physics at M.I.T., her computational soft matter research has been supported by the American Chemical Society's Petroleum Research Fund and the National Science Foundation's Division of Materials Research. She also studies computation in the physics curriculum, and race and gender issues in STEM. She is a Fellow of the American Physical Society, a member of its Committee on the Status of Women in Physics (CSWP), and on the Editorial Board of the American Journal of Physics. Her contact email is abug1@swarthmore.edu.

Adam Light is an Assistant Professor in the Department of Physics at Colorado College. Educated at Case Western Reserve University and The University of Colorado Boulder, he is interested in basic and applied plasma physics. Dr. Light is a member of the American Physical Society. His contact email is alight@coloradocollege.edu.